# Unstructured Inversion of New Hope

Ben Adler

**Abstract**

Introduced as a new protocol implemented in "Chrome Canary" for the Google Inc. Chrome browser, "New Hope" is engineered as a post-quantum key exchange for the TLS 1.2 protocol. The structure of the exchange is revised lattice-based cryptography. New Hope incorporates the key-encapsulation mechanism of Peikert which itself is a modified Ring-LWE scheme. The search space used to introduce the closest-vector problem is generated by an intersection of a tesseract and hexadecachoron, or the $\ell_\infty$-ball and $\ell_1$-ball respectively. This intersection results in the 24-cell $\mathcal{V}$ of lattice $\widetilde{\mathcal{D}_4}$. With respect to the density of the Voronoi cell $\mathcal{V}$, the proposed mitigation against backdoor attacks proposed by the authors of New Hope may not withstand such attempts if enabled by a quantum computer capable of implementing Grover's search algorithm.

## 1 Introduction

"New Hope" is a novel encryption scheme based on lattice cryptography and offers post-quantum security within the key exchange. New Hope uses a Montgomery form to reduce cost of implementation in terms of computational speed. New Hope reduces cost by sending an *x*-coordinate to compute the relative *x*-coordinate of any scalar [1]. Alkim, et al. implement a rounding function $\lfloor x \rceil$ derived from the work of Peikert [2] to achieve equality with the floor function $\lfloor x + 1/2 \rfloor$. This floor function is an element of integers.

New Hope employs $q = 12289$ and $n = 1024$ as constraints of lattice $D_4$, which results in a reduction of the modulus to $q = 12289 < 2^{14}$ [1]. Peikert defines both the rounding and floor function of New Hope, using $\delta$-subgaussian and zeta functions [2]. Peikert's "canonical embedding" necessarily incorporates a homomorphic injective ring that maps $(K)$ to $(\mathbb{C})$ which fixes pointwise$(\mathbb{Q})$ [2].

The critical nature of an unbiased modular operation presents key values which are assumed to mitigate cryptanalysis. Peikert recommends the use of small noise values to achieve this result while cautioning against cross-rounding given the determinancy that may result [2]. Any such determinancy negates an otherwise unbiased result. It is here that New Hope diverges from its basis on Peikert's work.

The creators of New Hope outline a sketch to create a backdoor in implementations of NTRU lattice-based cryptography. Concerns of a backdoor capability extended to New Hope will now be addressed in detail.

## 2 Parameters

The fixed parameter of $(a)$ may be a potential point of weakness against NTRU [1]. For mildly small values of $(f, g)$ where $f = g$, and $f = 1 \bmod p$ for some prime, $(p \geq 4 * 16 + 1)$ there is a point of weakness within the set

$$a = g\frac{1}{f} \bmod q$$

Insofar as $(a, b = as + e)$ it is possible to compute:

$$bf = afs + fe = gs + fe \bmod q$$

such that:

$$bf = gs + fe \bmod q$$

With small enough $(g, s, f, e)$, computing $gs + fe \in \mathbb{Z}$ once $(s \bmod q)$ is obtained proves the scheme is then corrupted. After establishing $t = s + e \bmod p$, with the coefficient of $(s)$ and $(e)$ smaller than$(16)$, the values of $(s, e)$ have sums within the range

$$\begin{cases} -2 * 16 \\ 2 * 16 \end{cases}$$

Knowing the values of $(s, e)$ within the range of $(-2 * 16, 2 * 16)$ in terms of:

$$\bmod p \ \geq 4 * 16 + 1$$

is knowing them in $\mathbb{Z}$. The manipulation to create a backdoor relies on the *pseudo-inverse* of a polynomial $(p)$ as the polynomial $(P \in \mathcal{P})$ such that $(P * p * s \equiv s \bmod q)$ for any polynomial $(s \in \mathcal{P})$ such that

$$s(1) \equiv 0 \bmod q$$

As long as the secret key equation can be modified to equal $t \equiv h * v + w \bmod q$ it is feasible to apply a pseudo-inversion. For a detailed analysis of Inversion Oracles refer to the primary source of Mol and Yung [3]. Through implementing the attack developed by Mol and Yung it is possible to show that an attacker possessing both a classical and quantum computer is capable of a backdoor attack against New Hope.

## 3 Inversion

Given the new secret key equation derived from [3], let the following hold:

$$(w = u - g), (v = F) \text{ for } t \equiv u - p_q * h (\bmod q)$$
$$(v = u - F), (w = g) \text{ for } t \equiv p_q * h + h * u (\bmod q)$$

In both cases, $(w, v)$ are binary. An oracle will output the correct key pair only when $\left(e \in E_{q,h}^{d_r}\right)$ [3]. To apply this inversion the anti-derivative of the Peikert scheme used by New Hope must be established. According to the authors of New Hope, the implementation of the key encapsulation method (KEM) relies on pseudorandom ring elements exchanged between Alice and Bob which are then used to derive the session key [1]. Alice then employs the ring element $(us = sas' + e's)$ and Bob uses$(v = bs' + e'' = sas' + es' + e'')$. The reconciliation function is defined as $\left(\text{rec}(w, b)\right)$ such that

$$\text{rec}(w, b) = \begin{cases} 0, \text{if } \left(w \in I_b + E (\text{mod } q)\right) \\ 1, \text{otherwise} \end{cases}$$

The authors of New Hope set as parameters of the polynomial ring,

$$\mathcal{R}_q = \frac{\mathbb{Z}_q[X]}{X^n + 1}$$

The message sent by Alice is denoted as$(b)$, while Bob's response is$(u, r)$ and an element of the ring $(R_q)$. The polynomial $(a \in \mathcal{R}_q)$ is public and constant. To generate the function which results in$(s \text{ mod } q)$, the algebraic manipulation itself is fairly straightforward. To begin deriving the necessary function to generate the secret key for an NTRU scheme, a pre-established value equal to $s \text{ mod } q$ is introduced:

$$as - s * \left(\frac{1}{a} - 1\right) = s \text{ mod } q$$

Via substitution, values of the variables $(a, b)$ already provided are used to calculate values of $t$.

$$(a, b) = as + e$$

$$as - s = b - t$$

After trivial algebraic manipulations, the values of $t$ can be equated to a set of equations, wherein the value of the constant $(a)$ can be substituted with previously afforded values given in [1].

$$t = \begin{cases} -as + s + b \\ s + e \text{ mod } p \end{cases}$$

Returning to the equations used to calculate $t$, new values of $t$ are now substituted and the two previous equations are calculated as equal to one another.

$$\left((as - s) * \left(\frac{1}{a} - 1\right) = (b - t) * \left(\frac{1}{a} - 1\right)\right)$$

Where $(a = fg^{-1} \bmod q)$ it is then possible to assert $(as - s + t = b)$, which in turn produces the primary equation for solving the value of $s \bmod q$.

By producing an equation that results in a required value for a backdoor attack against some NTRU lattice-based cryptography, the equation of $\left((as - s) * \left(\frac{1}{a-1}\right) = s \bmod q\right)$ generates the final steps to calculating the secret $(s)$. Using substitution yet again, but this time of the variable $a$, one derives:

$$\left((fg^{-1} \bmod q)s - s\right) * \left(\frac{1}{fg^{-1} \bmod q - 1}\right) = s \bmod q$$

By simplifying the equation, we then produce:

$$\frac{(fg^{-1} \bmod q)s - s}{(fg^{-1} \bmod q) - 1} = s \bmod q$$

By stating the division in an alternate form, one then has:

$$s = s \bmod q$$

The value of the variable $(q)$ is itself equivalent to $1 \bmod 2n$. Bearing in mind that $n = 1024$, it is known that $q \equiv 1 \bmod 2048$. An abbreviated integer table of equivalent values to $1 \bmod q$ is provided in Table 1.

| | | |
|---|---|---|
| 2049 | 4097 | 6145 |
| 8193 | 10241 | 12289 |
| 14337 | 16385 | 18433 |

*Table 1*

The anti-derivative, or indefinite integral pertinent to this analysis is defined by the variable $a$ which is equal to $fg^{-1} \bmod q$, which produces the equation:

$$\int \frac{\left(f\left(\frac{1}{g}\right) \bmod q\right) s - s}{\left(f\left(\frac{1}{g}\right) \bmod q\right) - 1} \mathrm{d}g = gs + \text{constant}$$

Returning to the exchange between Alice and Bob, Alice uses the equation $(us = sas' + e's)$ to send Bob a message, which Bob then uses the equation $(v = bs' + e'' = sas' + es' + e'')$ to reconcile the pair with. If the equation of

$$(s = s \bmod q)$$

can be shown to equal $\left(t \equiv h * v + w(\bmod q)\right)$, then an oracle output to break the encryption is feasible. The further constraint of $\left(e \in E_{q,h}^{d_r}\right)$ is also required. Returning to the values produced by$(t)$, let $(t)$ be equal to the following

$$t = \begin{cases} -as + s + b \\ s + e \bmod p \end{cases}$$

To satisfy the constraint of the variable $(e)$ as a member of $(E_{q,h}^{d_r})$ and with the value of $(q)$ known, one can substitute for $(e)$ accordingly. Where $(d_r)$ corresponds to the Hamming weights to produce an inversion oracle against NTRU [3], New Hope employs a weight value of $\left(\exp\left(\frac{-x^2}{2\sigma^2}\right)\right)$ to all integers $(x)$ such that there is no fixed value for $(a)$ [1], but rather each coefficient of $(a)$ is chosen uniformly at random from $\mathbb{Z}_q$. The discrete Gaussian distribution $(D_{\mathbb{Z},\sigma})$ is parametrized by the Gaussian parameter $(\sigma \in \mathbb{R})$ defined by the previously mentioned weight of all $(x)$. The values of $(\mathbb{Z}_q)$ for an integer $(q > 1)$ must be within the quotient ring $(\frac{\mathbb{Z}}{q\mathbb{Z}})$ such that $\mathcal{R} = \frac{\mathbb{Z}[X]}{X^n+1}$ is the ring of integer polynomials modulo $X^n + 1$ where each coefficient is reduced modulo $(q)$.

## 4 Algorithmic Geometry

With the intersection Voronoi $\mathcal{V}$ 24-cell treated as a convex polytope, the 16-cell $\ell_1$-ball is a simplicial polytope while the $\ell_\infty$-ball together with the 16-cell are the only regular Euclidean 4-space tessellations.

Given these parameters, the 24-cell constructed as a Voronoi tessellation having center at $D_4$ for any point $x$ is expressed as:

$$x_i \in \mathbb{Z}^4 : \sum_i x_i \equiv 0 \bmod 2$$

If, for any $x_i = s$ there is some point where $s(1) \equiv 0 \bmod q$, the introduction of an inversion oracle is then verified.

**4.1 Grover Inversion Parameters**

Grover's algorithm was engineered to optimize searches of sub-spaces computed with qubits, for a full detailing of this algorithm see [4]. With respect to generalized searches computed by Grover's algorithm, the coefficients as an evolution of probability amplitudes has an initial probability amplitude: $\binom{a_0}{b_0} = \begin{pmatrix} \left(\cos\frac{\pi}{6}\right) \\ \left(\sin\frac{\pi}{6}\right) \end{pmatrix}$

Given the risk of executing Grover's algorithm at the expense of performance costs, care must be placed in how many times the computation is performed. The variable of iterations, $(m)$ is treated in terms of rotations as well as number of correct solutions $(t)$ [4]. In terms of optimization and performance, it has been shown in [4] that the number of calls of the respective function is $O\left(\sqrt{N/2}\right)$. The rotations themselves operate according to the initial amplitude vector, whereby the representation of $M$ rotates the subspace amplitude vector through the originating angle twice based on $t$ [4]. The iteration continues until the angle of the amplitude representative of the correct subspace is approximately $\pi/2$ where each iteration of $M$ is relative to the initial angle:

$$\binom{a_m}{b_m} = R^m\theta \binom{a_0}{b_0}$$

for each $\binom{a_0}{b_0}$.

The implementation of Grover's algorithm is treated via reliance on a "needle-in-the-haystack" oracle discussed in [4]. This oracle function, termed $f$ is defined within $k \in \mathbb{Z}$, where $0 \leq k \leq 2^n - 1$. The further restrictions upon $k$ are such that $f(k) = 0$ for any $k$ with the single exception of $f(k_0) = k = 1$.

By virtue of the work conducted in [5], for any given $t$ which is unknown with respect to the number of solutions, $m$ is then treated as any arbitrary integer. Accordingly, $j$ is then any arbitrary integer within the range of the uniform distribution of $(0, m-1)$ [5]. As suggested by Boyer, et al. the extension of Grover's algorithm to implement Shor's algorithm as an additional process generates the ability to not only find solution $t$, but also count all solutions $t$. The resulting theorem from the work of Boyer, et al. is an examination of the number of solutions $t$ such that for disjoint subsets, where $T$ is the set of solutions and $N_t = \lfloor N/t \rfloor$ is the number of solutions. The resulting inequality, denoted as:

$N_t - \sqrt{N_t} - \sum_{X_i \in T} \left| \alpha'_{r,\overline{X_i}} \right| \leq 2r^2$. The result of the proof offered by [5] is the number of queries to the oracle to determine if $y \in A$ where $y$ is chosen arbitrarily requires at least $M$ iterations for $M = \left\lfloor \sin {}^{\pi}/_{8} \sqrt{N} - 1 \right\rfloor$.

## 5 Cryptanalysis

Treating the lattice $\widetilde{D_2}$ as a binary field extension of the approximate *x* coordinates, the binary field characteristic is thus two given the use of a Montgomery form for optimization [1]. This characteristic of two implies the binary field extension has order $2^n$ for *n*. Given $q = 12289$, this is equivalent to $q \equiv 1 \bmod 2n$ and any treatment of the Voronoi cell in terms of the reduced lattice must necessarily commute to the lattice in four dimensions. This occurs with any Boolean ring of characteristic two. To complete the inversion, we must now turn to an examination of the variable $e$.

We begin by treating the *e* over the range of *x*-coordinates. For $b = as + e$ we may easily derive $-e = \frac{as}{b}$. For characteristic two, any element is also it's additive inverse, thus $e = \frac{as}{b}$. Substituting the value of *e* for Voronoi coordinates *x*, we then find $x = \frac{as}{b}$ is equivalent to $x = \frac{(as+e)s}{as+e}$ as previously demonstrated. This trivially reduces to $x = s$, but more importantly results in $x - s = 0$. Using the property of additive inverse again, we then rephrase the equation as $-s - s = 0$. Thus, $-2s = 0$. Returning to mod $q$, as derived, we may reduce the equation $-2s = 0$ with respect to modulo $q$. The inversion constraints of $v = u - f$ and $w = g$ with respect to $t \equiv hv + w \bmod q$ for *(w,v)* are adjusted via substitution of *w* for *g* and *v* with $u - f$. We then return to the equivalent form of $s \bmod q$ which yields:

$$\frac{(fw^{-1} \bmod q)s - s}{(fw^{-1} \bmod q) - 1}$$

Using the pseudo-inverse polynomial $(P, p)$ we proceed using the expression $s \bmod q$ congruent to $Psp$. Allowing the polynomial *p* as an element of the ring $R_q$ and deriving *u* via substitution with respect to $v = u - f$ we may begin constructing the expression congruent to *t* by adding $p_q$ and *u*. We now must demonstrate recovery of $(s, e)$ with respect to the range mod $p \geq 4 * 16 + 1$. Having isolated $x$ equal to $s$ and then showing $-2s = 0$, we apply the additive inverse to produce $-s - s = s + s$.

We now have $s + s = 0 = s(1 + 1)$. For a coefficient of *x* resulting in $e \bmod p$, and with knowledge of public key *h*, we then compute $hv + w \bmod q$. Allowing $q = 2$ per the constraint of values $\mathbb{Z}_q, q > 1$ with respect to $f = 1 \bmod p$ let:

$$s = s \bmod q \equiv s \bmod 2$$

Using rec(*w,b*) as a function of *s*,

$$s(w){:}\, I_b + E(\bmod q)$$

and assume a zero is returned. The function *s(w)* for any instance in which the output is not zero results in a value of 1 being returned for $s(w)$ during reconciliation. Having shown $x = s$ and with knowledge of $b$ as values of $(a, s, e)$ we may substitute values as demonstrated in this work to recover:

$$s(1) = s = s \bmod q$$

$$=$$

$$s \bmod 2 \equiv 0 \bmod 2$$

## 6 Conclusion

With the values of $(e, s)$ available to an attacker, Oscar and leveraged against Alice, Oscar can maintain persistence in the network after inversion. New Hope largely avoids any issues this may cause by specifically using ephemeral $a$, but does allow temporary caching of $a$ with respect to Alice should the cost of generating unique $a$ become too great [1]. Considering this, it should be stated that if Oscar's inversion succeeds, his ability to access Alice's cached *a* hinges on knowing when the caching occurs, and what value of *a* is required.

The antiderivative provided opens the possibility to manipulate the secret *s* while simultaneously using the variable *g* substituted for *x* in addition to some constant. This constant added to the variables $(s, g)$ based upon traditionally fixed values of NTRU, though applied against New Hope facilitates an inversion of this scheme. The vectors of *x* and relative approximate coordinates are shown equivalent to *e* for known *s*. Under this model, Oscar can compromise the generation of $(e, s)$ and use these values to introduce an arbitrary constant.

References

[1] E. Alkim, L. Ducas, T. Poppelman and P. Schwabe, "Post-quantum key exchange - a new hope," International Association for Cryptologic Research, 2015.

[2] C. Peikert, "Lattice Cryptography for the Internet," in Post-Quantum Cryptography, Waterloo, Springer International Publishing, 2014, pp. 197-219.

[3] P. Mol and M. Yung, "Recovering NTRU Secret Key from Inversion Oracles," in Public Key Cryptography - PKC 2008, Barcelona, Springer Berlin Heidelberg, 2008, pp. 18-36.

[4] A. Pittenger, "An Introduction to Quantum Computing Algorithms" Birkhauser 2000

[5] M. Boyer, G. Brassard, P. Hoyer, A. Tapp "Tight Bounds on Quantum Searching," retrieved from https://arxiv.org/pdf/quant-ph/9605034.pdf on April 13, 2021